\documentclass[11pt]{article}

\usepackage{newpasp}
\usepackage{epsf}
\usepackage{graphicx}
\usepackage{epsfig}
\usepackage{rotating}
\usepackage{makeidx}

\def\edcomment#1{\iffalse\marginpar{\raggedright\sl#1\/}\else\relax\fi}
\reversemarginpar

\begin{document}
\title{The Interstellar Medium of Young Stellar Clusters from the Mid-Infrared Point of View}
 \author{Suzanne C. Madden}
\affil{CEA, Saclay, Service d'Astrophysique, Orme des Merisiers, 91191, Gif-sur-Yvette cedex, France, madden@discovery.saclay.cea.fr}
%\author{Ima Co-Author}
%\affil{The Name of My Institution, The Full Address of My Institution}

\begin{abstract}
Effects of young stellar clusters on their gas and dust environment
are probed using mid-infrared (MIR) wavelengths. The strong MIR
[NeIII]/[NeII] ratios ($\sim$ 5 to 10) reveal the presence of current
massive stars less than 5\,Myr.  Using MIR line ratios along with
optical and NIR data from the literature, composite SEDs are
constructed for NGC~1569, NGC~1140 and II\,Zw40. The stellar SEDs are
then used as input to a dust model to study the impact of the hard,
penetrating radiation field on the dust components, particularly in
low metallicity environments, where the destructive effects of the
massive stellar clusters on the environments occur on global
scales. For example, the smallest dust particles are destroyed over
larger regions in the dwarf galaxies than in normal metallicity
starbursts.
\end{abstract}


\section{Introduction}
The subject of stellar cluster formation history and environment has
made great headway lately, with the high resolution and sensitivity
currently available at optical and near infrared (NIR) wavelengths. In
principle, the mid-infrared (MIR) wavelength regime should provide
numerous advantages for such studies, since this wavelength range
is relatively extinction free (A$_{15\mu}$$_{m}$ $\sim$ 5\% A$_{J}$)
and contains diagnostic ionic lines to probe HII regions. In addition, hot
dust emission provides us with another link to the ultraviolet starlight 
that has been absorbed and reemitted by the nearby grains. Our knowledge of the
MIR wavelength window has been limited by the low spatial and spectral
resolution provided by the IRAS satellite, and has remained rather sketchy 
when it comes to detailed studies of the ISM of individual galaxies. 
The Infrared Space Observatory (ISO; Kessler et al. 1996) has been a recent
turning point in this effort, providing high spectral and spatial
resolution and unprecedented sensitivity in the MIR through the far
infrared (FIR).
We have incorporated these MIR and FIR observations in a study of the
energy redistribution in starburst galaxies, with the aim of
understanding the impact of the star formation on the surrounding gas and dust.

The main limitation in MIR star cluster studies remains the
spatial resolution, despite the great improvement over previous
instrumentation provided by ISOCAM ($\sim$ 6$\arcsec$ at
15$\mu$m; Cesarsky, C.J. et al. 1996). ISOCAM resolves about 600 pc at 20 Mpc, 
the distance of the closest massive merging system, the Antennae. 
However, even with this
limitation, we are able to draw noteworthy conclusions from the MIR,
from unique MIR diagnostics. Here, I concentrate primarily on results
of the nearer dwarf galaxies, since impacts of the massive
clusters on the global dust and gas environment are very pronounced
in these relatively small objects.

\section{What Do MIR Wavelengths Trace?}
Figure 1 shows ISOCAM 5-17\,$\mu$m spectra for 
the three dwarf galaxies II\,Zw40, NGC\,1140 and NGC\,1569, with
metallicities of {\hbox{$\,^1\!/_7$}} to {\hbox{$\,^1\!/_3$}}\,Z$_{\odot}$
(Madden et al. 2000), along with the spectrum
of the notoriously metal poor SBS\,0335-052 
({\hbox{$\,^1\!/_{40}$}}\,Z$_{\odot}$; Thuan, Sauvage, \& Madden 1999).  
All these spectra show obvious MIR signatures
of massive stars. 

As often seen in starburst
galaxies, the MIR spectra are dominated by steeply rising continua
longward of $\sim$ 10 $\mu$m. Thermal emission from hot small grains with
mean temperatures of the order of hundreds of Kelvin are responsible for the
MIR continuum emission. 

The unidentified infrared bands (UIBs) at 6.2,
7.7, 8.6, 11.3 and 12.6 $\mu$m, have been attributed to aromatic
hydrocarbon particles undergoing stochastic temperature fluctuations
(i.e, PAHs: L\'eger \& Puget 1984; Allamendola, Tielens, \& Barker
1989; coal grains: Papoular, Reynaud, \& Nenner 1991). They are observed
to peak in the photodissociation (PDR) zones around HII regions, 
but are destroyed deep within the HII regions themselves
(Verstraete et al. 1996; Cesarsky, D. et al. 1996; Tran 1998). 
While the UIBs are not obvious in the spectra of II\,Zw40 and
SBS\,0335-052, and are only very weakly present NGC~1569, they can be
distinguished in the spectrum of NGC~1140.  Several ground state
fine structure nebular lines are present also in 3 of the spectra, the
most prominent being 15.6 $\mu$m [NeIII] (ionisation potential $\sim$ 41
eV) and 10.5 $\mu$m [SIV] ($\sim$ 35 eV).  Weaker,
lower energy lines may also present, such as the 8.9 $\mu$m [ArIII]
line and the [NeII] 12.8 $\mu$m line, which can be blended with the
12.6 $\mu$m UIB.  

While all of these spectra are very different from one
another, all differ significantly from those of normal metallicity
starburst galaxies. Normal starburst galaxies show prominent UIBs, in
contrast to AGNs, which are devoid of UIBs (e.g. Roche et al. 1991;
Dudley 1999, Laurent et al. 2000; Sturm et al. 2000). When compared to
spectra characteristic of PDRs and HII regions (e.g. M17, Cesarsky, D. et
al. 1996; Verstraete et al. 1996), II\,Zw40 is remarkably similar to
that of an HII region. In contrast, NGC\,1140, which has a very flat
continuum yet a very strong [NeIII] line, does have a more obvious
contribution from PDR regions in its spectra. Note that the MIR spectrum of N66,
the most prominent HII region in the SMC, also shows a scarcity of
UIBs in the vicinity of the most massive central cluster (Contursi et
al. 2000), as does the low metallicity source NGC~5253 (Crowther et
al. 1999).

\begin{figure}[h]
\plotfiddle{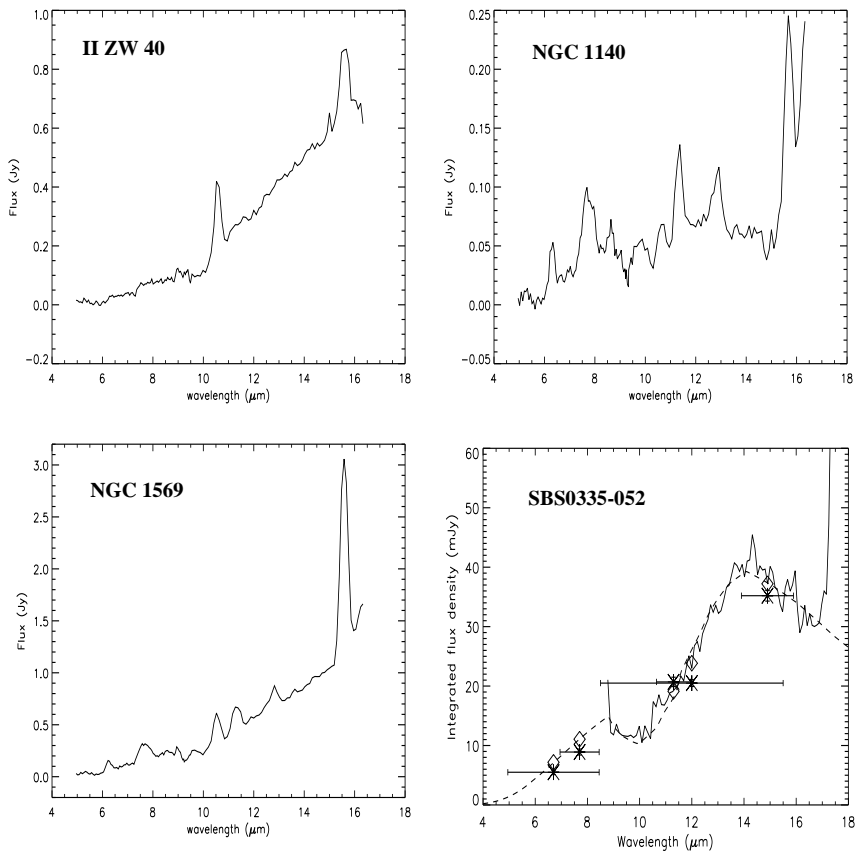}{11.0cm}{0}{52}{52}{-180}{-10}
\caption{MIR ISOCAM spectra of the dwarf galaxies II\,Zw40, NGC~1569,
NGC~1140 and SBS\,0335-052. The horizontal lines for SBS\,0335-052 are
broad band measurements; the dashed line is a blackbody with
A$_{v}$ $\sim$ 20 (from Thuan, Sauvage \& Madden 1999).
% 3 names kept to avoid ambiguity in list
Note the absorption
at $\sim$ 9 and 18 $\mu$m in SBS\,0335-052, attributed to amorphous silicates.}
\label{cvfs}
\end{figure}

In some starburst galaxies, amorphous silicate is seen in absorption
centered at 9 and 18 $\mu$m (Roche et al. 1991; Dudley 1999;
Laurent et al. 2000).  We can fit the MIR region of the II\,Zw40
spectrum with a blackbody at 193 K and an absorption equivalent to
A$_{v}$ $\sim$ 4. Dust temperatures derived assuming blackbodies
should be interpreted with care, since the dust emitting in the MIR is
expected to be undergoing stochastic processes rather than being
in thermal equilibrium with the radiation field. The amount of
absorption in II\,Zw40 (A$_{v}$ $\sim$ 4) has yet to be confirmed. In
SBS\,0335-052, A$_{v}$ $\sim$ 20 has been deduced from the absorption in 
the ISOCAM MIR spectra (Fig.\,1).  The presence of a
significant amount of dust at a metallicity as low as Z$_{\odot}$/40 is
surprising, especially since star formation in SBS\,0335-052 began as 
recently as 100 Myr ago (Papaderos et al. 1998; Thuan, Izotov, \& Foltz
1999). Such high extinction implies that the current star formation
rate, hidden by dust, might be underestimated by at least 50\% (Thuan,
Sauvage \& Madden 1999).
%3 names kept to avoid ambiguity in list.

\section{Effects of the Massive Star Formation on the Gas}
As a consequence of the smaller dust abundance of most dwarf galaxies, the
ISM throughout these galaxies is affected by the hard radiation
field of the massive stellar clusters. All star forming dwarf galaxies 
contain evidence for Wolf-Rayet stars (Schaerer, Contini, \& Pindao 1999) 
and super star clusters have been detected in NGC~1140 (Hunter, O'Connell, 
\& Gallagher 1994), NGC~1569 (O'Connell, Gallagher, \& Hunter 1994) 
and SBS\,0335-052 (Thuan, Izotov, \& Lipovetsky 1997). Their harsh 
radiation fields, which more readily permeate the ISM compared those in solar 
metallicity environments, are capable of destroying the UIB
carriers, for example, over very extensive spatial areas.  
The effect of the pervasive radiation field can be witnessed in NGC\,1569 
(Fig.\,2), where photodissociation occurs on global scales. 
Violent activity is revealed by the H$\alpha$ distribution 
(Waller 1991; Martin 1998) and the 15.8 $\mu$m [NeIII] emission,
with giant streamers suspected to originate from the energetic winds
of the super star clusters A \& B, (black stars in Fig.\,2).
The UIB, [SIV] and [NeIII] emission seems to avoid the super
star clusters, which blow out much of the gas and dust on relatively
short time scales. This effect is also seen in the CO (Taylor et
al. 1999), HI (Israel \& van Driel 1990) and H$\alpha$ (Waller
1991) distributions. Likewise we see the destruction of the UIBs in the
beam-averaged spectrum of the entire galaxies II\,Zw40 and SBS\,0335-052
(the available spatial resolution prevents us from seeing more detail
within these galaxies in the MIR).

\begin{figure}[h]
\begin{center}
%\plotone{ngc1569_halpha_ne3_xfig_bw.eps} 
%\plotfiddle{madden_fig2.eps}{7.0cm}{0}{60}{60}{-175}{-135} 
\includegraphics[clip=,width=9cm]{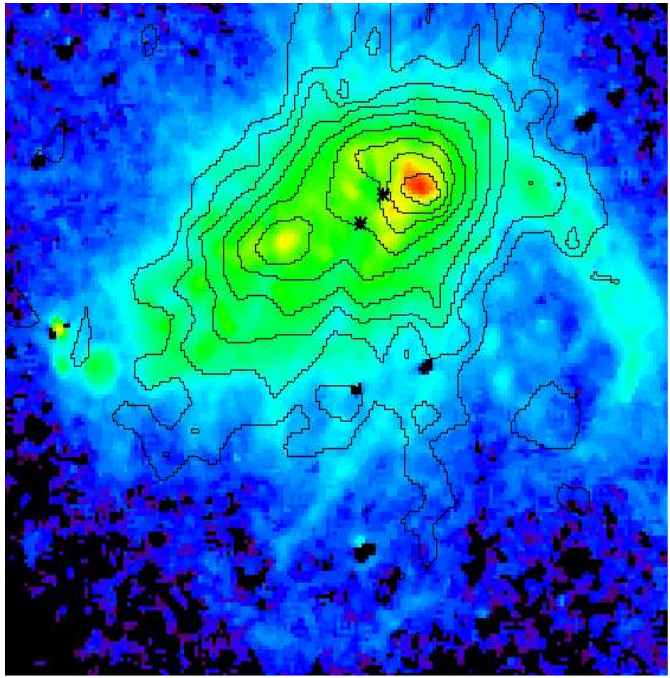}
\caption{NGC~1569: H$\alpha$ (image) (Waller 1991) and 15.8 $\mu$m [NeIII] 
emission contours. Note the extended [NeIII] filaments, also seen in H$\alpha$.
The 2 black stars mark the positions of the super star clusters A and B 
(O'Connell et al. 1994) which are devoid of [NeIII] emission.}
\label{n1569_image}
\end{center}
\end{figure}

\section{Modelling the Spectral Energy Distribution}
We have compiled broad band data from the literature for II\,Zw40,
NGC\,1569 and NGC\,1140 and have combined these with our MIR data to
construct appropriate stellar spectral energy distributions (SEDs). In
so doing, we fit the observed optical and NIR data with population
synthesis models of PEGASE (Fioc \& Rocca-Volmerange 1997), taking
into account the constraints of the MIR line emission by modelling the
corresponding photoionisation with CLOUDY (Ferland 1996).  After
briefly describing the results of this process, we discuss the results
of the use of the reconstructed stellar SEDs as input to our dust
model.

\subsection{Combined stellar evolution and photoionisation model results} 

When assuming instantaneous star formation, a metallicity Z$_{\odot}$/5
and a Salpeter IMF (with upper and lower mass cut-offs of 0.1
and 120 solar masses), we find solutions to the observed broad band
colours for a variety of ages and ionisation parameters. 
The ISOCAM MIR observations 
provide the diagnostic lines of neon, sulphur and argon, that have been
recently addressed e.g. by Lutz et al. (1998), Crowther
et al. (1999), Schaerer \& Stasi\'nska (1999) and Genzel et
al. (1998).  

For example, the [NeIII]/[NeII] ratio is a measure of
T$_{eff}$, the hardness of the radiation field, and therefore traces
the massive stellar population.  For the dwarf galaxies, we find
[NeIII]/[NeII] ratios in the range of 5 to 10 - much higher values
than those of normal metallicity galaxies ($\leq$1; Thornley et al. 2000). The
extreme values of the [NeIII]/[NeII] ratios are related to the
low metallicities of the systems: the T$_{eff}$ of the stars increases
as the metallicity decreases for a specific stellar age.  High ratios
of [NeIII]/[NeII] and the prominent [SIV] in these spectra limit the
age of the present star formation to $<$ 5 Myr.  Beyond this age, the
massive stars have died and the [NeIII]/[NeII] ratio drops
dramatically. 
The high excitation 24.9 $\mu$m [OIV] line, covered by
the ISO SWS data, is observed in some dwarf galaxies (Lutz et
al. 1998) and has been attributed to the presence of
Wolf-Rayet stars (Schaerer \& Stasi\'nska 1999). 

For NGC\,1569, NGC\,1140 and II\,Zw40, we construct composite stellar SEDs 
that require 70 to 95 \% of the stellar mass to be provided by an 'older' 
population with ages between about 10 and 30 Myr, with
the remaining 5\% to 30\% corresponding to a very young population
($<$\,5\,Myr). Observational evidence for the presence of
Wolf-Rayet stars corroborates the existence of this very young stellar 
population (Vacca \& Conti 1992). The broad band optical and NIR data 
alone reveal predominantly the older population in our apertures. 
Fig.\,3 shows an example of the resultant composite SED for II\,Zw~40, 
including the extreme ultraviolet (EUV) radiation that the young, 
massive stellar population traces.

\begin{figure}[h]
%\includegraphics[width=\textwidth, viewport=0 0 670 495, clip]{sed_zw1b.eps}
%\plotone{sed_zw1b.eps}
%\plotfiddle{madden_fig3.eps}{8cm}{0}{50}{50}{-180}{-20} 
\centerline{\includegraphics[clip=,width=11cm]{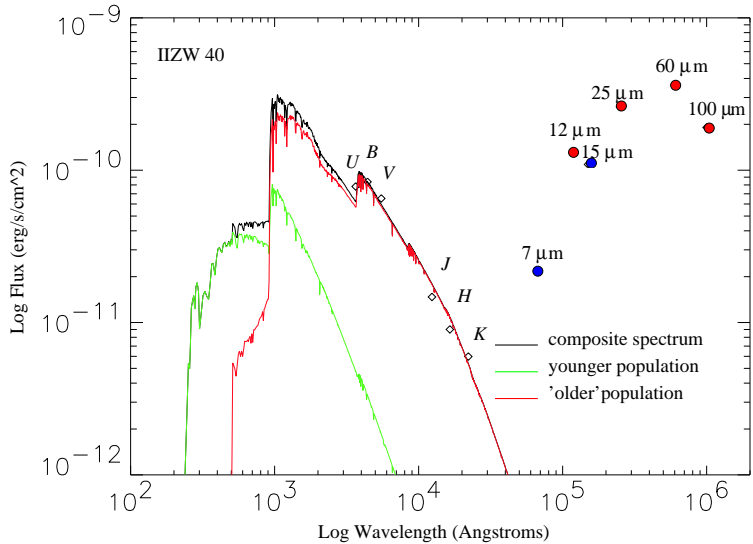}}
\caption[]{II\,Zw40 SED. The synthetic stellar spectra are fits to the 
extinction-corrected optical and NIR data from the literature for a 12'' 
aperture using PEGASE. The 12, 25, 60 and 100 $\mu$m data are from IRAS 
and the 7 and 15 $\mu$m data points are integrated over 5.0 to 8.5 $\mu$m 
and 12.0 to 17 $\mu$m bands, respectively, using the ISOCAM spectrum (Fig.\,1). 
A composite SED is shown, with a 5\% contribution in mass from a young 
population ($\sim$ 3 Myr) and 95\% from an ``older'' (10 Myr) population, 
which provides most of the observed optical and NIR fluxes.}
\label{sed}
\end{figure}

\subsection{Effects of the Massive Star Formation on the Dust}

We use the modelled stellar spectra of II\,Zw40, NGC~1569 and NGC~1140 as
input to a dust model to study the effects of this radiation field on
the dust properties. This is an important step since dust plays a
major role in influencing the chemical and physical state of the
ISM. We use the D\'esert, Boulanger, \& Puget (1990) model to fit the
various dust components emitting in the MIR and the FIR. This model
calculates the IR emission from large silicate grains (BGs), very
small amorphous carbon grains (VSGs), and stochastically heated
polycyclic aromatic hydrocarbons (PAHs), for various grain size
distributions. 

In these three galaxies the MIR spectrum is clearly 
dominated by emission from VSGs with very little PAH emission. 
The BG component dominates the overall dust emission with mass fractions 
ranging from 93\% to 99\%, while the PAH mass fraction is
relatively insignificant --- 5 orders of magnitude lower. 
The model gives a PAH/VSG mass ratio of $2-3\,10^{-4}$ 
for NGC~1569 and II\,Zw 40 and 10 times this value for NGC~1140. 
For comparison, the D\'esert et al. model applied to the Galactic cirrus 
gives a PAH/VSG mass ratio $\sim 1$. 
Thus, even compared to the VSG population, we find an
insignificant mass fraction of PAHs, reflecting the fact that the PAHs
are destroyed throughout the entire galaxies, as a result of the hard
radiation fields originating from the few massive stellar
clusters. This in an important result, since PAHs are thought to
be the primary particles responsible for the photoelectric heating
process (Bakes \& Tielens 1994) and are incorporated in PDR models
(Kaufman et al. 1999). Our preliminary results, while not
statistically robust at this stage, suggest that even in the absence
of PAHs, the photoelectric effect is efficient, as both II\,Zw40 and
NGC~1569 are relatively prominent [CII] sources among the 
galaxies surveyed (Jones et al. 1997). 
On the contrary, in NGC\,1140, where PAHs are more
obvious in the MIR spectra (Fig.\,1), we do not detect [CII]. VSGs
with derived sizes of $\sim$40 to 300\,\AA, which are very abundant 
relative to the PAHs in NGC~1569 and II\,Zw40 but less so in
NGC~1140, may therefore be the more efficient sources of
photoelectric gas heating in these environments, rather than
PAHs. 

This scenario is in contrast to normal metallicity galaxies,
where the PAHs are prominently observed, and the effects of the
numberous massive stellar activity are much more local rather than global.

\section{Summary}
MIR ISOCAM spectroscopy provides details of ionic lines, UIBs and
the distribution of small hot grain emission in dwarf galaxies. The strong
MIR [NeIII]/[NeII] ratios are signatures of the hard radiation fields
and indicate the presence of clusters of young massive stars in
dwarf galaxies. Because of the increase in T$_{eff}$ in low
metallicity environments, this ratio is enhanced in dwarf galaxies
to at least 5 to 10 times that observed in normal metallicity galaxies. The
penetrating radiation field also affects the dust components,
destroying the UIBs in some dwarf galaxies on global scales, as is
evident in the MIR spectra and in the dust modeling. This dramatic
global effect of the massive stellar population in dwarf galaxies, due
to the decrease in attenuation of the UV flux, is not apparant in
normal metallicity galaxies, where these effects are experienced much more
locally.

\acknowledgements
It is a pleasure to acknowledge my collaborators D. Ragaigne  and A. Jones. 
I wish to also thank W. Waller for his H$\alpha$ image of NGC\,1569.

\end{document}